\documentclass[nofootinbib]{nature}

\usepackage{amsmath}    
\usepackage{color}
\usepackage[10pt]{moresize}

\usepackage{graphicx}
\usepackage{caption}
\DeclareCaptionLabelFormat{adja-page}{\hrulefill\\#1 #2 \emph{(previous page)}}
\usepackage{subfig}

\usepackage{amsfonts}
\usepackage{amssymb}
\usepackage{amscd}
\usepackage{amsmath}
\usepackage{enumerate}
\usepackage{epsfig}
\usepackage{epstopdf}
\usepackage{dcolumn}
\usepackage{bm}
\usepackage{amsthm}
\usepackage{amsfonts}
\usepackage{color} 
\usepackage[usenames,dvipsnames]{xcolor}
\usepackage{amsmath}
\usepackage{enumerate}
\usepackage{bbm}
\usepackage[colorlinks=true,citecolor=blue,urlcolor=black]{hyperref}

\usepackage{mathrsfs}

\begin{document}

\title{Single-photon computational 3D imaging at 45 km}

\author{Zheng-Ping Li$^{1,2,\ast}$,
Xin Huang$^{1,2,\ast}$,
Yuan Cao$^{1,2,\ast}$,
Bin Wang$^{1,2}$,
Yu-Huai Li$^{1,2}$,
Weijie Jin$^{1,2}$
Chao Yu$^{1,2}$,
Jun Zhang$^{1,2}$,
Qiang Zhang$^{1,2}$,
Cheng-Zhi Peng$^{1,2}$,
Feihu Xu$^{1,2,\dag}$,
Jian-Wei Pan$^{1,2,\dag}$
}

\maketitle

\begin{affiliations}
\item Shanghai Branch, National Laboratory for Physical Sciences at Microscale and Department of Modern Physics, University of Science and Technology of China, Shanghai 201315, China.
 \item CAS Center for Excellence and Synergetic Innovation Center in Quantum Information and Quantum Physics,
University of Science and Technology of China, Shanghai 201315 China.
\\ $^\ast$These authors contributed equally to the paper.
 \\ $^\dag$e-mail: feihu.xu@ustc.edu.cn; pan@ustc.edu.cn.
\end{affiliations}

\maketitle

\begin{abstract}
Long-range active imaging has a variety of applications in remote sensing and target recognition. Single-photon LiDAR (light detection and ranging) offers single-photon sensitivity and picosecond timing resolution, which is desirable for high-precision three-dimensional (3D) imaging over long distances. Despite important progress, further extending the imaging range presents enormous challenges because only weak echo photons return and are mixed with strong noise. Herein, we tackled these challenges by constructing a high-efficiency, low-noise confocal single-photon LiDAR system, and developing a long-range-tailored computational algorithm that provides high photon efficiency and super-resolution in the transverse domain. Using this technique, we experimentally demonstrated active single-photon 3D-imaging at a distance of up to 45 km in an urban environment, with a low return-signal level of $\sim$1 photon per pixel. Our system is feasible for imaging at a few hundreds of kilometers by refining the setup, and thus represents a significant milestone towards rapid, low-power, and high-resolution LiDAR over extra-long ranges.
\end{abstract}

Long-range active optical imaging has widespread applications, ranging from remote sensing\cite{marino2005jigsaw,schwarz2010lidar,glennie2013geodetic}, satellite-based global topography\cite{smith1998topography,abdalati2010icesat}, and airborne surveillance\cite{glennie2013geodetic}, to target recognition and identification\cite{gschwendtner2000development}. An increasing demand for these applications has resulted in the development of smaller, lighter, lower-power LiDAR systems, which can provide high-resolution three-dimensional (3D) imag¬ing over long ranges with all-time  capability. Time-correlated single-photon-counting (TCSPC) LiDAR is a candidate technology that has the potential to meet these challenging requirements\cite{buller2007ranging}. Particularly, single-photon detectors\cite{hadfield2009single} and arrays\cite{richardson2009low,villa2014cmos} can provide extraordinary single-photon sensitivity and better timing resolution than analog optical detectors\cite{buller2007ranging}. Such high sensitivity allows lower-power laser sources to be used and can permit time-of-flight imaging over significantly longer ranges. Tremendous efforts have thus been devoted to the development of single-photon LiDAR for long-range 3D imaging\cite{buller2007ranging,mccarthy2009long,mccarthy2013kilometer,li2017multi}. Single-photon 3D imaging up to a distance of 10 km has been reported\cite{pawlikowska2017single}.

In long-range 3D imaging, a frontier question is the distance limit; i.e., over what distances can the imaging system work? For a single-photon LiDAR system, the echo light signal, and thus the signal-to-noise ratio (SNR), decreases rapidly with imaging distance $R$, which imposes limits on the image quality. Overall, for a given system with laser power $(P)$ and telescope aperture $(A)$, the limit of operational range $R_{\text{limit}}$  is determined by the SNR and the algorithm efficiency as follows:
\begin{equation} \label{eq1}
\begin{aligned}
R_{\text{limit}} \propto \text{SNR}\cdot\eta_{a} = \frac{PA\eta_{s}}{R^2h\nu\bar{n}}\cdot\eta_{a}.
\end{aligned}
\end{equation}
Here the free parameters, $\eta_{s}$, $\bar{n}$, $\eta_{a}$ are the system-detection efficiency, background noise, and reconstruction-algorithm efficiency, respectively. The free parameters $\eta_{s}$ and $\bar{n}$ are determined by the hardware design, whereas $\eta_{a}$ is determined using the algorithm design, in which the computational approach can greatly increase its efficiency\cite{kirmani2013first}.

Recent developments in active imaging have become increasingly dependent on computational power\cite{altmann2018quantum}. Computational optical imaging, in particular, has seen remarkable progress\cite{velten2012recovering,sun20133d,gariepy2015detection,o2018confocal,xu2018revealing,saunders2019computational,shin2015photon,altmann2016lidar,shin2016photon,rapp2017few,lindell2018single}. An important research trend today is the development of efficient algorithms for imaging with a small number of photons\cite{shin2015photon,altmann2016lidar,shin2016photon,rapp2017few,lindell2018single,morris2015imaging}. One advance in this area was the demonstration of high-quality 3D structure and reflectivity in the laboratory environment by an active imager detecting only one photon per pixel (PPP), based on the approaches of pseudo-array\cite{shin2015photon,altmann2016lidar}, single-photon camera\cite{shin2016photon}, unmixing signal/noise\cite{rapp2017few} and machine learning\cite{lindell2018single}. These algorithms have the potential to greatly improve the imaging range.

\begin{figure}[htbp]
\centering\includegraphics[width=15.7cm]{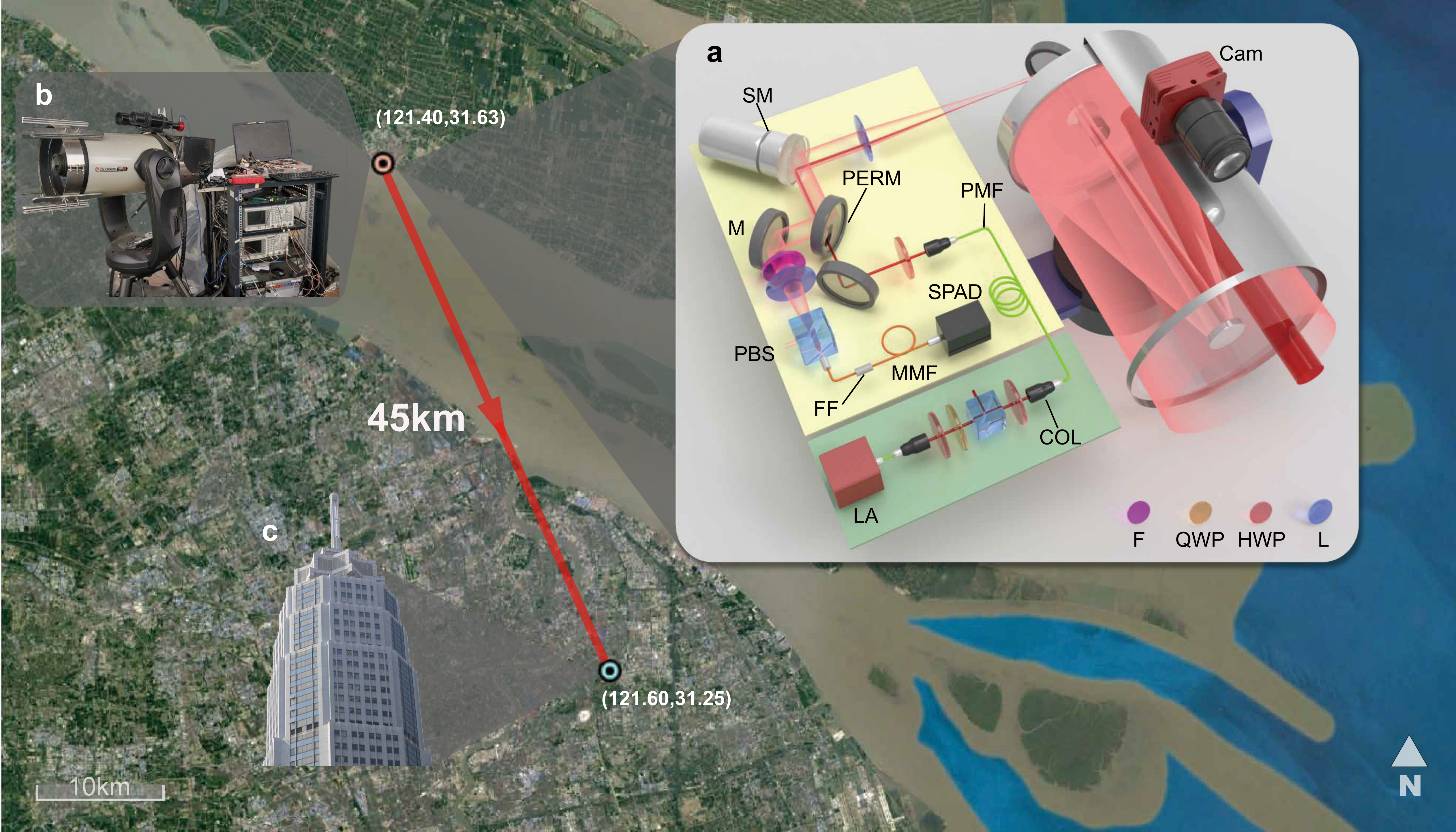}
\caption{\textbf{Illustration of long-range active single-photon LiDAR.} Satellite image of the ex-periment layout in the urban area of Shanghai City, with the single-photon LiDAR positioned on Chongming Island. \textbf{a}, Schematic diagram of experimental setup. SM, scanning mirror; Cam, camera (visible band); M, mirror; PERM, $45^\circ\ $ perforated mirror; PBS, polarization beam splitter; SPAD, single-photon avalanche diode detector; MMF, multimode fiber; PMF, polarization-maintaining fiber; LA, laser (1550 nm); F, filters (long pass and 9 nm bandpass); FF, fiber filter (1.3 nm bandpass); L, lens; HWP, half-wave plate; QWP, quarter-wave plate; EDFA, erbium-doped fiber amplifier. \textbf{b}, Photograph of experimental setup, including the optical system (left) and the electronic control system (right). The optical system consists of a telescope congregation  and an optical-component box for shielding. \textbf{c}, Close-up photograph of the target, the Pudong Civil Aviation Building, which is on the opposite shore of the river from Chongming Island. The building is 45 km from the single-photon LiDAR setup.}
\label{Fig:experiment}
\end{figure}

Our primary interest in this work is to significantly push the imaging range towards the ultimate limit for high-resolution 3D imaging. We approached this problem by developing an advanced technique based on \emph{both} hardware and software implementations that are specifically designed for long-range single-photon LiDAR. On the hardware side, we designed a high-efficiency coaxial-scanning system (see Fig.~\ref{Fig:experiment}) to more efficiently collect the weak echo photons and more strongly suppress system background noise. On the software side, we developed a computational algorithm that offers super-resolution in transverse domain and high photon efficiency for the data of low photon counts (i.e., $\sim$1 signal PPP) mixed with high background noise (i.e., SNR $\sim$1/30). These improvements allow us to demonstrate super-resolution single-photon 3D imaging over a distance of 45 km from $\sim$1 signal PPP in an urban environment.

\textbf{Setup.} Figure~\ref{Fig:experiment} shows a bird's-eye view of the long-range active-imaging experiment, which was set up at Chongming Island in Shanghai city facing a target building located across the river. The optical transceiver system incorporated a commercial Cassegrain telescope with a 280 mm aperture and a high-precision two-axis au¬tomatic rotating stage to allow large-scale scanning of the far-field target. The optical components were assembled on a custom-built aluminum platform integrated with the telescope tube. The entire optical hardware system is compact and suitable for mobile applications (see Fig.~\ref{Fig:experiment}b).

Specifically, as shown in Fig.~\ref{Fig:experiment}a, a standard erbium-doped near-infrared fiber laser (1550 nm, 500 ps pulse width, 100 kHz repetition rate) served as the light source for illumination. Operating in the near-infrared range makes the system eye-safe, reduces solar background, has low atmospheric loss, and is compatible with telecom optical components. The maximal average laser power used was 120 mW. The laser output was vertically polarized and was coupled into the telescope through a small aperture consisting of an oblique hole through the mirror. The beam was expanded and output with a divergence angle of about 35 $\mu rad$. The transmitted and received beams were coaxial, allowing the area illuminated by the beam and the field of view (FoV) to remain matched while scanning. The returned photons were reflected by the perforated mirror and collected by a focal lens. A polarization beam splitter (PBS) served to couple only the horizontally polarized light into a multimode fiber. Finally, the photons were spectrally filtered, coupled efficiently into a multimode fiber, and detected by an InGaAs/InP SPAD (single-photon avalanche diode) detector\cite{Yu2017Fully}. Further details of the setup are given in the Supplementary Information.

To achieve a high-efficiency, low-noise confocal single-photon LiDAR system, we implemented several optimized designs, most of which differed from those used in previous experiments\cite{mccarthy2009long,mccarthy2013kilometer,li2017multi,pawlikowska2017single}. First, we developed a two-stage FoV scanning method---offering both fine-FoV and wide-FoV scanning---to simultaneously maintain fine features and expand the entire FoV. For fine-FoV scanning, we used a scanning mirror mounted on a piezo tip-tilt platform to steer the beam in both \emph{x} and \emph{y} axial directions. This coplanar dual-axis scanning scheme is capable of high-precision angle scanning with highly simplified optical elements, thereby avoiding imaging pillow distortions. For wide-FoV scanning, we used a two-axis automatic rotation table to rotate the entire telescope. Next, the inter-pixel spacing was chosen to match the FoV of half a detector pixel. This strategy gives high image resolution and allows us to achieve super-resolution combined with our computational algorithm (see below). In addition, we used the polarization degree of freedom to reduce the internal back-reflected background noise, which was achieved by using orthogonally polarized inputs and outputs. Finally, we used miniaturized optical holders to align the apertures of all optical elements to a height of 4 cm, thereby improving the system stability. The entire optical platform was compact, measuring only $30\times30\times35\ cm^3$, including a customized aluminum box to block the ambient light, and was mounted behind the telescope.

\textbf{Algorithm.} The long-range operation of the LiDAR system involves two important challenges that limit the image reconstruction: (i) The diffraction limit and turbulence in the outdoor environment leads to a large FoV in the far field that covers multiple reflectors with different depths (see Supplementary Fig. S3), which greatly deteriorates the image resolution. (ii) The extremely low SNR limits the unmixing of signal from noise in an optical beam. These two challenges are unique to the long-range operation and were thus not considered by previous computational algorithims\cite{kirmani2013first,shin2015photon,altmann2016lidar,shin2016photon,rapp2017few,lindell2018single}. In particular, the widely assumed condition\cite{rapp2017few,lindell2018single} ``one depth per pixel" is not valid for long-range operation.

We developed a photon-efficient super-resolution algorithm to solve these two challenges. The forward model of the imaging system is shown in Methods. The implementation of the algorithm may be divided into two steps: (1) We developed a \emph{global gating} approach to unmix signal from the noise. In this approach, we sum the detection counts from all the pixels to form a time-domain histogram, and then do a peak search to extract the time bins corresponding to signal detections (see Supplementary Fig. S1). The key idea is that natural scenes have a finite number of reflectors that are clustered in depth (i.e., time) and therefore can be effectively filtered out in the time domain. (2) We constructed a modified SPIRAL-TAP solver\cite{harmany2010spiral} to directly solve the inverse problem with a \emph{3D matrix}, which differs from previous algorithims\cite{kirmani2013first,shin2015photon,altmann2016lidar,shin2016photon,rapp2017few,lindell2018single} that were implemented on a two-dimensional 2D matrix. For long-range measurements, detection at each pixel involves a convolution operation with two kernels; one in the spatial domain (within the FoV) and one in the temporal domain. We recorded the measurements from all pixels in a 3D matrix to maintain the features of reflectivity and depth and, to solve the deconvolution problem, we use the total-variation norm to do a direct convex optimization on this 3D matrix with a transverse-smoothness constraint. In this way, the system provides \emph{super-resolved} reconstructions of reflectivity and depth (see Supplementary Information).

\textbf{Results.} We present an in-depth study of our imaging system and algorithm for a variety of targets with different spatial distributions and structures over different ranges. The experiments were done in an urban environment. Depth maps of the targets were reconstructed by using the proposed algorithm with $\sim$1 PPP for signal photons and a SNR as low as 0.03, where the SNR is defined for a time gate of 200 ns (corresponding to an image depth of 30 m). We also made accurate laser-ranging measurements to determine the distance to the targets (see Supplementary Information).

\begin{figure}[!t]\center
\resizebox{16.3cm}{!}{\includegraphics{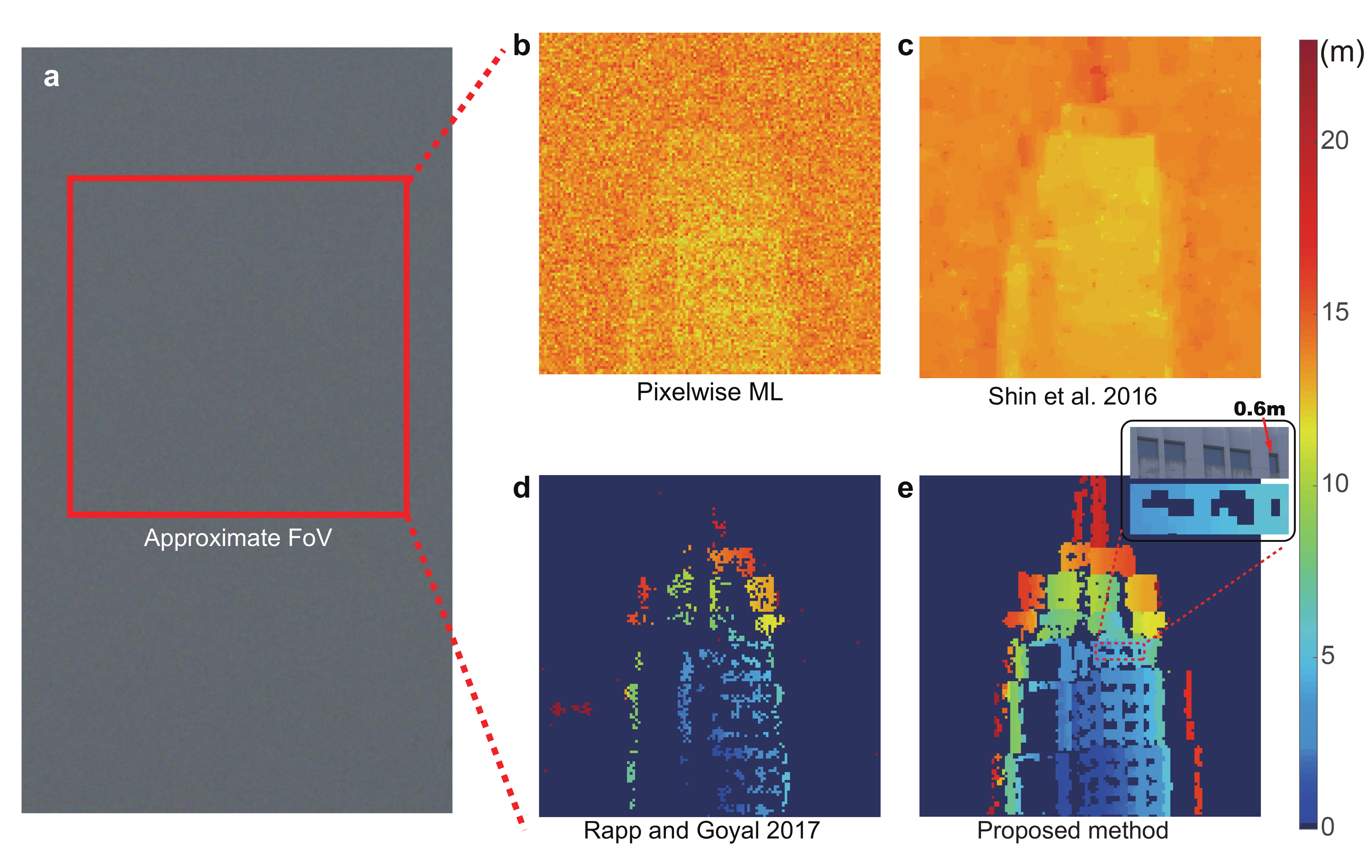}}

\caption{\textbf{Long range 3D imaging over 45 km.} \textbf{a}, Real visible-band image (tailored) of the target taken with a standard astronomical camera. This photograph is substantially blurred due to the inadequate spatial resolution and the air turbulence in the urban environment. The red rectangle indicates the approximate LiDAR FoV. \textbf{b--e}, The reconstruction results obtained by using the pixelwise maximum likelihood (ML) method, the photon-efficient algorithm by Shin \emph{et al.}\cite{shin2015photon}, the unmixing algorithm by Rapp and Goyal\cite{rapp2017few}, and the proposed algorithm, respectively. The single-photon LiDAR recorded an average PPP of $\sim$2.59, and the SNR was $\sim$0.03. The calculated relative depth for each individual pixel is given by the false color (see color scale on right). Our algorithm performs much better than the other state-of-art photon-efficient computational algorithms and provides super-resolution sufficient to clearly resolve the 0.6-m-wide windows (see expanded view in inset of panel \textbf{e}).}
\label{Fig:45km}
\end{figure}

We first show the imaging results for a long-range target, called the Pudong Civil Aviation Building, at a one-way distance of 45 km. Fig.~\ref{Fig:experiment} shows the topology of the experiment. The imaging setup was placed on the 20th floor of a building and the target was on the opposite shore of the river. The ground truth of the target is shown in Fig.~\ref{Fig:experiment}c. Fig.~\ref{Fig:45km}a shows a visible-band photograph, taken with a standard astronomical camera (ASI120MC-S). This photograph cannot show any shape of the target due to the inadequate spatial resolution and the air turbulence in the urban environment. We adopted our single-photon LiDAR to do the imaging at night and produce a $(128\times128)$-pixel image. A modest laser power of 120 mW was used for the data acquisition. The averaged PPP was $\sim$2.59, and the SNR was 0.03. The plots in Fig.~\ref{Fig:45km}b-e show the reconstructed depth obtained by using various imaging algorithms, including the pixelwise maximum likelihood (ML), the photon-efficient algorithm by Shin \emph{et al.}\cite{shin2015photon}, the unmixing algorithm by Rapp and Goyal\cite{rapp2017few}, and the algorithm proposed herein. The proposed algorithm recovers the fine features of the building, allowing the scenes with multilayer distribution to be accurately identified. The other algorithms, however, fail in this regard. These results clearly demonstrate clearly that the proposed algorithm operates remarkably well for spatial and depth reconstruction of long-range targets. More importantly, by fine-interval scanning (half FoV spacing), the proposed algorithm achieves a transverse resolution of 0.6 m, which resolves the small windows of the target building (see inset in Fig.~\ref{Fig:45km}e). This resolution overcomes the transverse diffraction limit of the single-photon LiDAR system, which is about 1.0 m at the far field of 45 km (see Supplementary Information).

\begin{figure}[!t]\center
\resizebox{16cm}{!}{\includegraphics{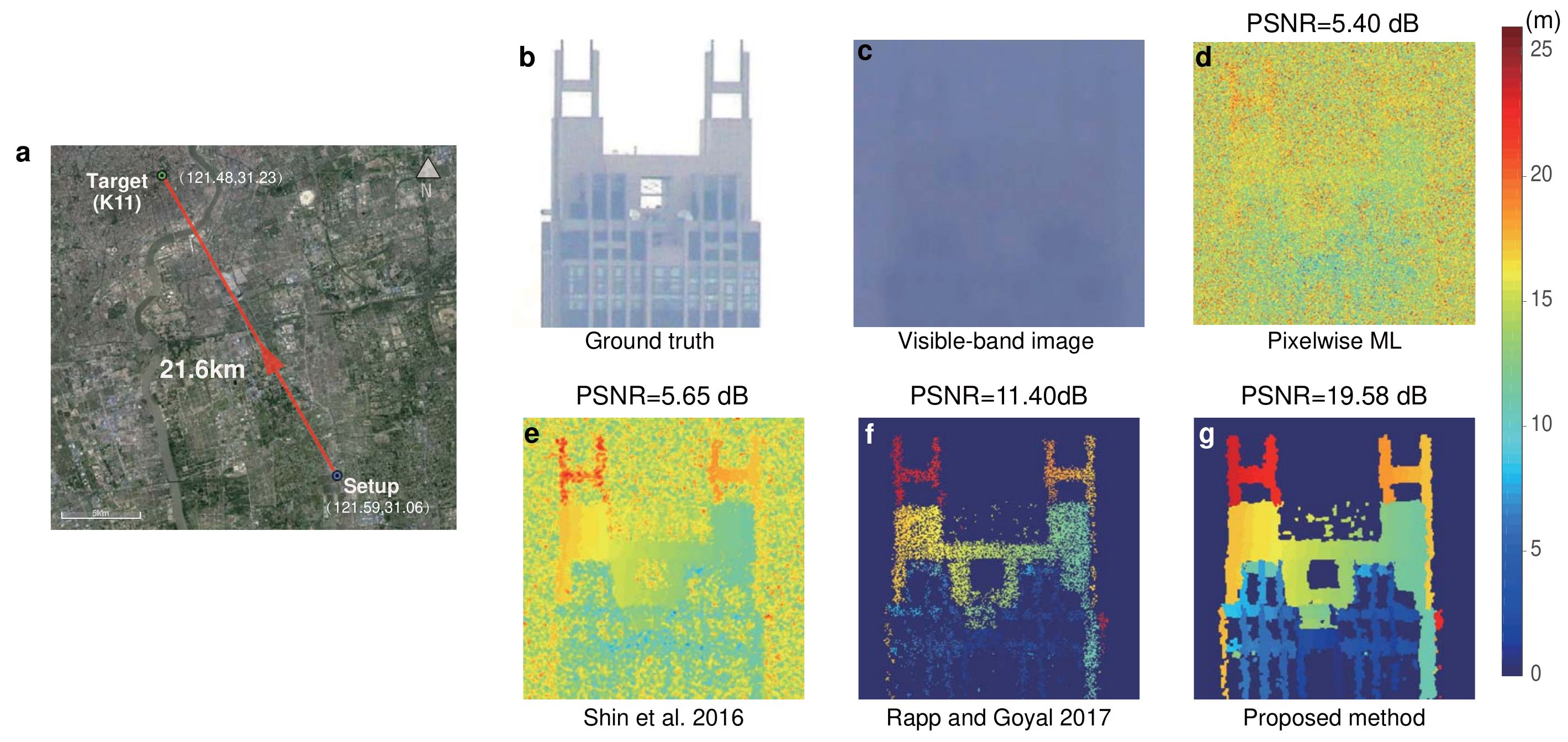}}
\caption{\textbf{Long-range target taken in daylight over 21.6 km.} \textbf{a}, Topology of the experiment. \textbf{b},  Ground-truth image of the target (building K11). \textbf{c}, Visible-band image of the target taken with a standard astronomical camera. \textbf{d-g}, Depth profile taken with the proposed single-photon LiDAR in daylight and reconstructed by applying the different algorithms to the data with 1.2 signal PPP and SNR = 0.11. \textbf{d}, Reconstruction with the pixelwise ML method. \textbf{e}, Reconstruction with the algorithm of Shin \emph{et al.}\cite{shin2015photon}. \textbf{f}, Reconstruction with the algorithm of Rapp and Goyal\cite{rapp2017few}. \textbf{g}, Reconstruction with the proposed algorithm. The peak signal to noise ratio (PSNR) was calculated by comparing the reconstructed image with a high-quality image obtained with a large number of photons. The proposed method yields a much higher PSNR than the other algorithms.}
\label{Fig:k11}
\end{figure}

To quantify the performance of the proposed technique, we show an example of a 3D image obtained in daylight of a solid target with complex structures at a one-way distance of 21.6 km  (see Fig.~\ref{Fig:k11}a). The target is a part of a skyscraper called K11 (see Fig.~\ref{Fig:k11}b) that is located in the center of Shanghai city. Before data acquisition, a photograph of the target was taken with a visible-band camera (see Fig.~\ref{Fig:k11}c); the resulting visible-band image is blurred because of the long object distance and the urban air turbulence. The single-photon LiDAR data were acquired by scanning $256\times256$ points at an acquisition time per point of 22 ms and with a laser power of 100 mW. The average PPP was 1.20, and the SNR was 0.11. The plots in Fig.~\ref{Fig:k11}d-g show the reconstructed depth profiles using the pixelwise ML method, the photon-efficient algorithm\cite{shin2015photon}, the unmixing algorithm\cite{rapp2017few}, and the proposed algorithm herein. The proposed algorithm allows us to clearly identify the shape of the grid structure on the walls, the symmetrical H-like structure at the top of the building, and the small left-to-right gradient caused by the oblique perspective. The quality of the reconstruction is quantified based on the peak signal to noise ratio (PSNR) by comparing the reconstructed image with a high-quality image obtained by using a large number of photons. The PSNR of the proposed algorithm is 14 dB better than that of the ML method, and 8 dB better than that of the unmixing algorithm.

\begin{figure}[!t]\center
\resizebox{15.7cm}{!}{\includegraphics{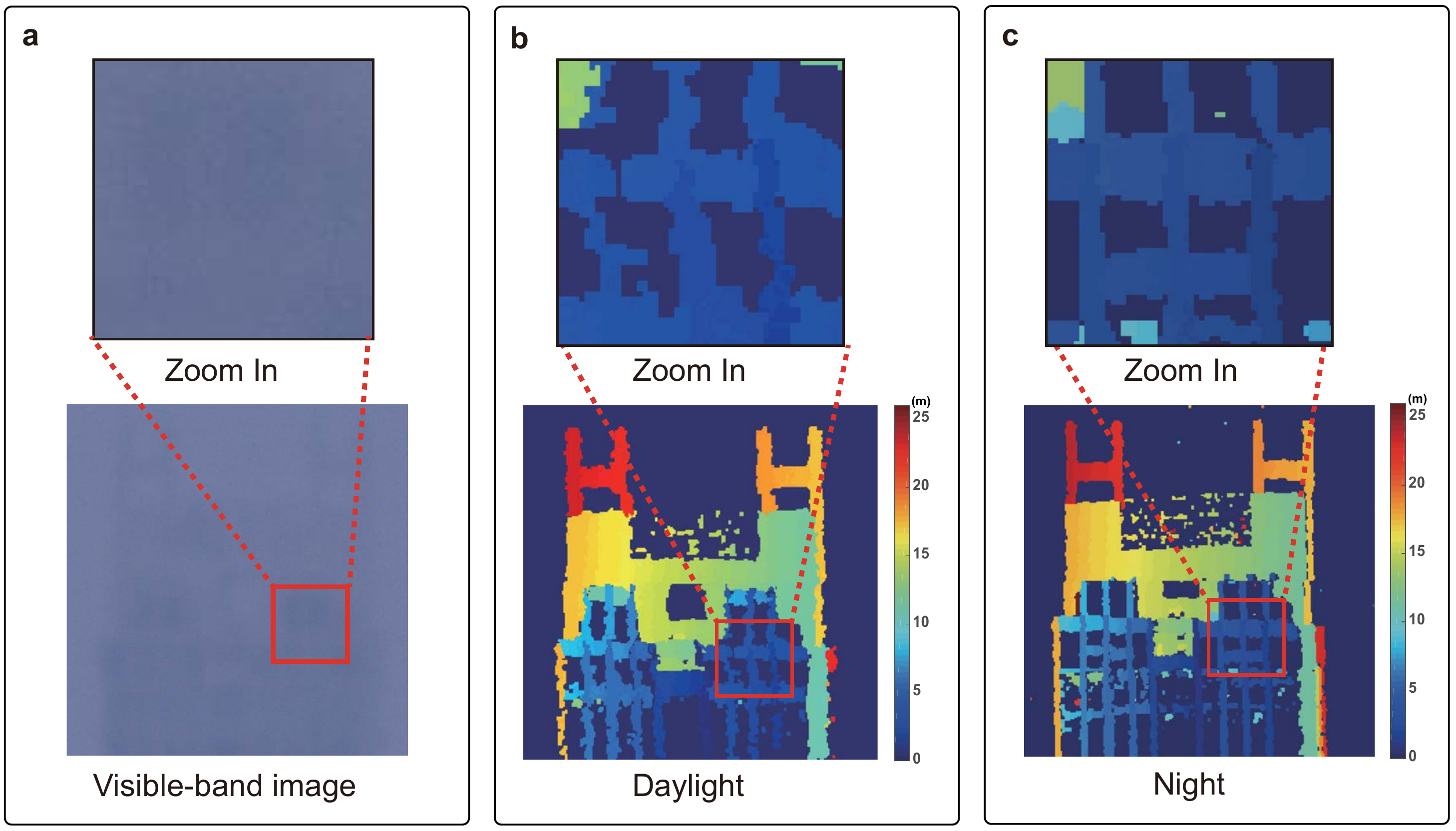}}
\caption{\textbf{Long-range target at 21.6 km imaged in daylight and at night.} \textbf{a}, Visible-band image of the target taken with a standard astronomical camera. \textbf{b}, Depth profile of image taken in daylight and reconstructed with signal PPP=1.2, SNR=0.11. \textbf{c}, Depth profile of image taken at night and reconstructed with signal PPP=1.2, SNR=0.15.}
\label{Fig:multi-layer2}
\end{figure}

To demonstrate the all-time capability of the proposed LiDAR system, we used it to image building K11 both in daylight and at night (i.e., 11:00 AM and 12:00 PM) on June 15, 2018 and compared the resulting reconstructions. The proposed single-photon LiDAR gave 1.2 signal PPP and a SNR of 0.11 (0.15) in daylight (at night). Fig.~\ref{Fig:multi-layer2}b and Fig.~\ref{Fig:multi-layer2}c show front-view depth plots of the reconstructed scene. The single-photon LiDAR allows the surface features of the multilayer walls of the building to be clearly identified both in daylight and at night. The enlarged images in Fig.~\ref{Fig:multi-layer2}b and Fig.~\ref{Fig:multi-layer2}c show the detailed features of the window frames, although, due to increased air turbulence during the day, the daytime image is slightly blurred compared with the nighttime image.

\begin{figure}[!t]\center
\resizebox{14.5cm}{!}{\includegraphics{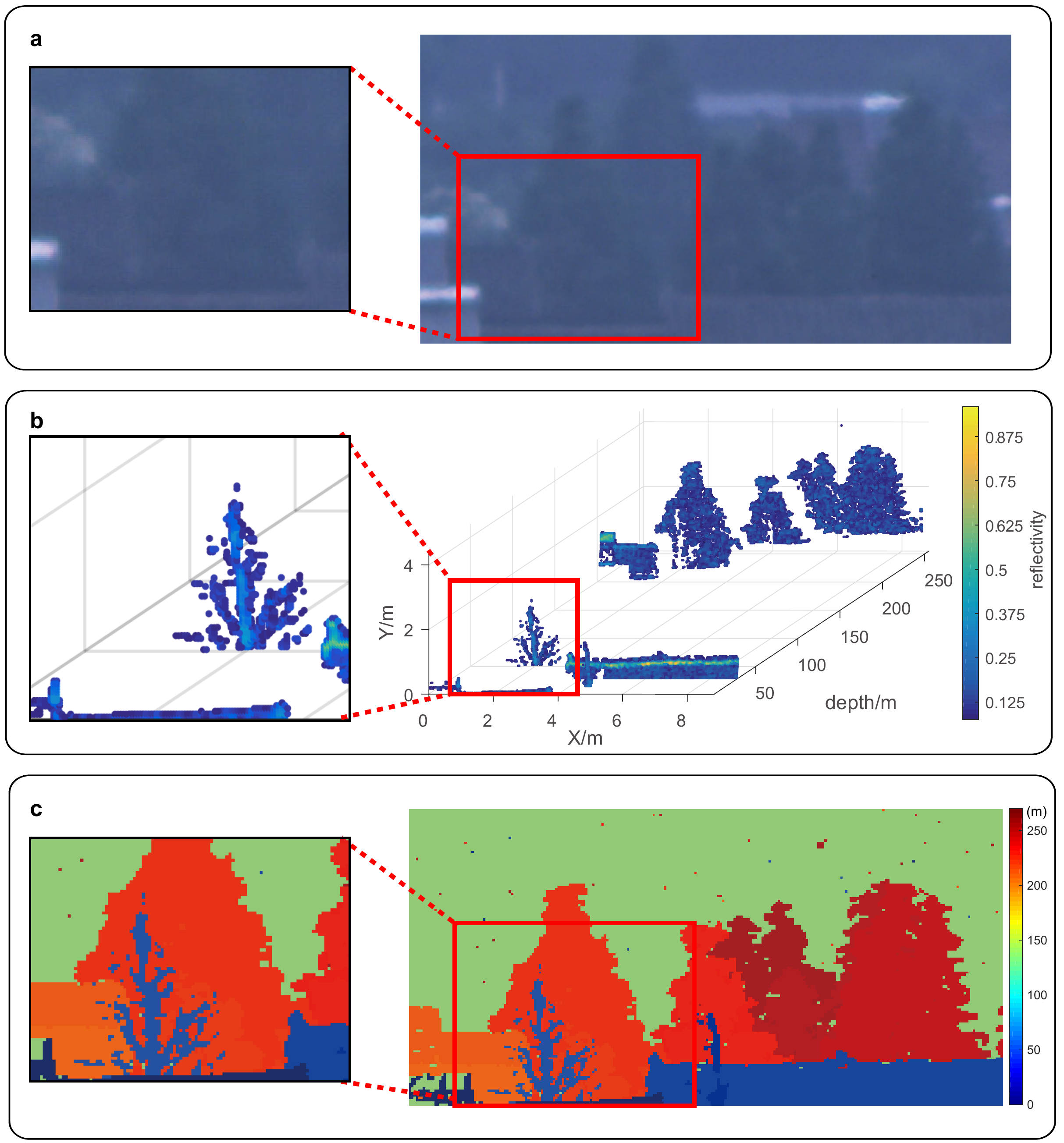}}
\caption{\textbf{Reconstruction of multilayer depth profile of a complex scene.} \textbf{a}, Visible-band image of the target taken by a standard astronomical camera mounted on the imaging system with an f = 700 mm camera lens. \textbf{b,c,} Depth profile taken by the proposed single-photon LiDAR over 2.1 km, and recovered by using the proposed computational algorithm. Trees at different depths and their fine features can be identified.}
\label{Fig:multi-layer}
\end{figure}

Finally, Fig.~\ref{Fig:multi-layer} shows a more complex natural scene with multiple trees and buildings at a one-way distance of 2.1 km. This scene were selected and scanned in daytime to produce a $(128\times256)$-pixel depth image. Fig.~\ref{Fig:multi-layer}b shows the depth profiles of the scene, and Fig.~\ref{Fig:multi-layer}c shows a depth-intensity plot. The conventional visible-band photograph in Fig.~\ref{Fig:multi-layer}a is blurred mainly because of smog in Shanghai, and does not resolve the different layers of trees in the 2D image. In contrast, as shown in Fig.~\ref{Fig:multi-layer}b,c, the proposed LiDAR system clearly resolves the details of the scene, such as the fine features of the trees. More importantly, the 3D capability of the single-photon LiDAR system clearly resolves the multiple layers of trees and buildings (see Fig.~\ref{Fig:multi-layer}b). This result demonstrates the superior capability of the near-infrared single-photon LiDAR system to resolve targets through smog\cite{tobin2019three}.


To summarize, we demonstrate in this work active single-photon 3D imaging at ranges of up to 45 km. The 3D images are generated at the single-photon-per-pixel level and allow for target recognition and identification at very low light levels. The proposed high-efficiency confocal single-photon LiDAR system, noise-suppression method, and advanced computational algorithm opens new opportunities for rapid and low-power LiDAR imaging over long ranges. These results should facilitate the adaptation of the system for use in future single-photon LiDAR systems with Geiger-mode detector arrays\cite{richardson2009low,villa2014cmos,shin2016photon} for rapid data acquisition of moving targets or for fast imaging from moving platforms. By refining the setup, our system is feasible for a few hundreds of kilometers (see Supplementary Information). Overall, our results open a new venue for high-resolution, fast, low-power 3D optical imaging over ultralong ranges.


\section*{Acknowledgments}

This work was supported by National Key R\&D Program of China (SQ2018YFB050100), National Natural Science Foundation of China, the Chinese Academy of Science, the Thousand Young Talent Program of China, the Fundamental Research Funds for the Central Universities and the Shanghai Science and Technology Development Funds (18JC1414700). The authors would like to thank Cheng Wu, Ting Zeng, and Qi Shen for helpful discussions.

\section*{Author Contributions}

All authors contributed extensively to the work presented in this paper.

\section*{Author Information}
The authors declare no competing financial interests.

\section*{Methods}

\subsection{Forward model.}

For long-range imaging, the diffraction limit of telescope projects each detector pixel to a large spot (FoV) that covers multiple points in the far field. The rate function for the scanning angle $(\theta_x,\theta_y)$ is the convolution of the depth-reflectivity map and a spatial-temporal kernel,
\begin{equation} \label{eq2}
\begin{aligned}
R(t;\theta_x,\theta_y)=\int_{\theta'_x,\theta'_y\in \text{FoV}}h_{xy}(\theta_x-\theta'_x,\theta_y-\theta'_y)r(\theta'_x,\theta'_y)h_t(t-2d(\theta'_x,\theta'_y)/c)d\theta'_xd\theta'_y+b,
\end{aligned}
\end{equation}
where FoV denotes the field -of -view of the scanner, [$r(\theta'_x,\theta'_y)$,$d(\theta'_x,\theta'_y)$] is the [intensity,depth] pair, $c$ is the speed of light, and $b$ is the background rate, and $h_{xy}$ and $h_t$ are spatial and temporal kernels representing the intensity distribution of the FoV and the shape of the laser pulse.

From the theory of photodetection, the total photons detected for all pixels follow a Poisson distribution, which can be represented by a 3D matrix as
\begin{equation} \label{eq3}
\begin{aligned}
\mathbf{S}\sim \text{Poisson}(\mathbf{h}\ast \mathbf{RD}+\mathbf{B}).
\end{aligned}
\end{equation}

Here, $\mathbf{h}_{xy}$ and $\mathbf{RD}_{xy}$ are discrete representations of the function $h_{xy}h_t$ and $r(\theta'_x,\theta'_y)\delta(t-2d(\theta'_x,\theta'_y)/c)$ respectively, $\ast$ is the convolution operation, and $\mathbf{B}$ is the 3D matrix of background noise. The goal of image reconstruction is to estimate $\mathbf{RD}$, which contains intensity and depth information, from the low-resolution photon-histogram data $\mathbf{S}$.

\subsection{Algorithm.}

Previous state-of-the-art photon-efficient algorithms\cite{kirmani2013first,shin2015photon,altmann2016lidar,shin2016photon,rapp2017few,lindell2018single} cannot be applied to long-range imaging because the common assumption\cite{rapp2017few} of ``one depth per pixel" made in these studies is not valid for Eq.~\eqref{eq1}. We thus developed in this work a computational algorithm tailored specifically for long-range 3D imaging. The implementation of this algorithm may be divided into the following two steps: \textbf{(1)} We developed a global gating approach to unmix signal from noise. We sum the detection counts from all the pixels to form a histogram in the time domain, and then apply a peak-searching procedure to extract the time bins for signal detection. \textbf{(2)} We solve the deconvolution problem by using a modified SPIRAL-TAP solver\cite{harmany2010spiral}, where we generalized the solver from a 2D matrix to a 3D matrix. Specifically, with $\mathscr{L}_{RD}$  being the negative log-likelihood function of Eq.~\eqref{eq3}, the proposed algorithm solves the following problem:
\begin{equation} \label{eq4}
\begin{aligned}
&& \mathop{minimize}\limits_{\mathbf{RD}}\   \Phi \triangleq \mathscr{L}_{\mathbf{RD}}(\mathbf{RD}; \mathbf S, \mathbf h, \mathbf B )+\beta_{TV}\ \text{pen}_{TV}(\mathbf{RD}) \\
&& subject \ to \ \ \mathbf{RD}_{i,j,k}\geq 0, \forall i,j,k.
\end{aligned}
\end{equation}
Here we impose a transverse-smoothness constraint by using the total-variation (TV) norm to exploit the spatial correlations of natural scenes. After minimization, a depth map is constructed by calculating the average time of arrival for each pixel in the 3D matrix. Further details of the proposed computational algorithm are given in the Supplementary Information.

\clearpage

\end{document}